\documentclass[preprint,showpacs,prb,preprintnumbers,amsmath,amssymb,]{revtex4}

\usepackage{graphicx}
\usepackage{dcolumn}
\usepackage{bm}
\begin{document}

\title{Chemical control of the charge state of nitrogen-vacancy centers in diamond}

\author{M.V. Hauf}
\affiliation{Walter Schottky Institut, Technische Universit\"{a}t
M\"{u}nchen, Am Coulombwall 4, 85748 Garching, Germany}
\author{B. Grotz}
\author{B. Naydenov}
\affiliation{3rd Physics Institute and Research Center SCoPE, University of Stuttgart, Pfaffenwaldring 57, 70550 Stuttgart, Germany}
\author{M. Dankerl}
\affiliation{Walter Schottky Institut, Technische Universit\"{a}t
M\"{u}nchen, Am Coulombwall 4, 85748 Garching, Germany}
\author{S. Pezzagna}
\affiliation{RUBION, Ruhr-Universit\"{a}t Bochum, Universit\"{a}tsstrasse 150, 44780 Bochum, Germany}
\affiliation{Research Department IS 3/HTM, Ruhr-University Bochum, 44780 Bochum, Germany}
\author{J. Meijer}
\affiliation{RUBION, Ruhr-Universit\"{a}t Bochum, Universit\"{a}tsstrasse 150, 44780 Bochum, Germany}
\author{F. Jelezko}
\author{J. Wrachtrup}
\affiliation{3rd Physics Institute and Research Center SCoPE, University of Stuttgart, Pfaffenwaldring 57, 70550 Stuttgart, Germany}
\author{M. Stutzmann}
\affiliation{Walter Schottky Institut, Technische Universit\"{a}t
M\"{u}nchen, Am Coulombwall 4, 85748 Garching, Germany}
\author{F. Reinhard}
\email{f.reinhard@physik.uni-stuttgart.de}
\affiliation{3rd Physics Institute and Research Center SCoPE, University of Stuttgart, Pfaffenwaldring 57, 70550 Stuttgart, Germany}
\author{J.A. Garrido}
\email{JoseAntonio.Garrido@wsi.tum.de}
\affiliation{Walter Schottky Institut, Technische Universit\"{a}t
M\"{u}nchen, Am Coulombwall 4, 85748 Garching, Germany}

\date{\today}

\begin{abstract}
We investigate the effect of surface termination on the charge state of nitrogen vacancy centers, which have been ion-implanted few nanometers below the surface of diamond. We find that, when changing the surface termination from oxygen to hydrogen, previously stable NV$^-$ centers convert into NV$^0$ and, subsequently, into an unknown non-fluorescent state. This effect is found to depend strongly on the implantation dose. Simulations of the electronic band structure confirm the dissappearance of NV$^-$ in the vicinity of the hydrogen-terminated surface. The band bending, which induces a p-type surface conductive layer leads to a depletion of electrons in the nitrogen-vacancies close to the surface. Therefore, hydrogen surface termination provides a chemical way for the control of the charge state of nitrogen-vacancy centers in diamond. Furthermore, it opens the way to an electrostatic control of the charge state with the use of an external gate electrode.
\end{abstract}

\pacs{73.20.At, 78.20.-e, 78.55.-m, 76.30.Mi}

\maketitle

Nitrogen-vacancies (NV) in the diamond lattice have been extensively studied in the past for several reasons. NV centers can act as single photon emitters in the visible range with absolute photo stability at room temperature \cite{Kurtsiefer2000}. They have found applications in novel fields like quantum computation \cite{Neumann2010} and single spin magnetometry \cite{Maze2008, Balasubramanian2008,Balasubramanian2009} due to the possibility to prepare and read out their corresponding spin state optically \cite{Jelezko2004} and because of long spin coherence times (T2 = 1.8ms @RT) \cite{Balasubramanian2009}. They occur in several charged states, where the neutral (NV$^0$) and the negatively charged (NV$^-$) state are the most common ones. NV$^-$ is the prevailing state for centers buried deep in high-purity material. However, it has been observed to become unstable and to turn into NV$^0$ close to the surface of diamonds \cite{Fu2010} as well as in nanodiamonds \cite{Bradac2010, Rondin2010}. For applications such as magnetometry the control of the NV charge state is a crucial point, as defects with reliable spin properties in close vicinity to the surface are required. Therefore, it is of great interest to understand the influence of the surface \cite{Santori2009}  and gain control over the charge state of the NV centers in diamond. \\

\begin{figure}
	\centering
		\includegraphics{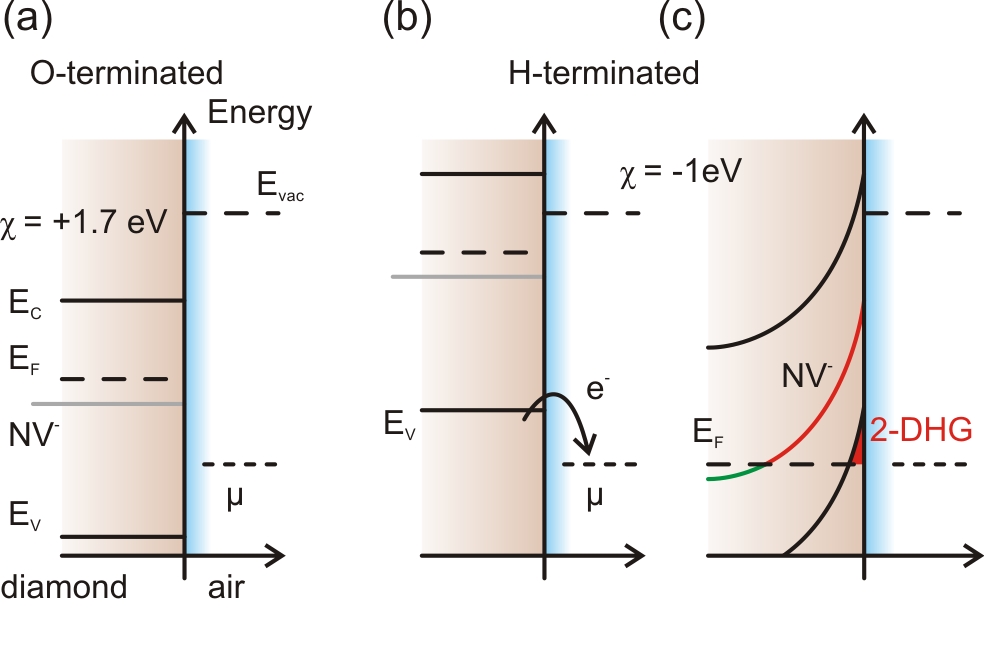}
		\caption{Energy band schematic of diamond. (a) O-terminated diamond: The conduction band $E_C$ lies below the vacuum level $ E_{vac}$ (Electron affinity $\chi = +1.7$ eV). The NV$^-$ level lies beneath the Fermi level $E_F$, where the position is mainly determined by the bulk concentration of nitrogen. (b) H-terminated diamond: The bands are shifted upwards by the hydrogen termination, leading to a negative electron affinity ($\chi \approx -1.0$ eV ). Electrons can transfer into acceptor states $\mu$ of an adsorbed water layer. (c) Equilibrium is established and a two-dimensional hole gas (2-DHG) is induced at the surface. Close to the surface, the strong band bending modifies the occupancy probability of the impurity levels such that the NV$^-$ level lies above $E_F$.}
		\label{fig:fig2}
\end{figure}
The properties of diamond surfaces are known to depend strongly on their surface termination. For instance, in an oxygen-terminated diamond, the band-gap of 5.45 eV results in a negligible concentration of free charge carriers [Fig. \ref{fig:fig2}(a)], given that no impurities are present in the diamond crystal. On the other hand, an accumulation of holes is observed when the diamond surface is hydrogenated \cite{Landstrass1989}. The C-H bonds induce an effective surface dipole moment which shifts the conduction and valence band upwards, leading to a negative electron affinity $\chi \approx -1$ eV [Fig. \ref{fig:fig2}(b)] \cite{Garrido2008a}. If the diamond is brought in contact with air, a layer of atmospheric adsorbates will form at its surface. These adsorbates provide energy levels which serve as acceptor states for electrons from the diamond valence band and defects in the band gap. Electrons are transfered into the adsorbate layer until the Fermi-level $E_F$ in the diamond has equilibrated with the electrochemical potential $\mu$ in the adsorbate layer\cite{Maier2000}. This leads to a band bending at the diamond surface and the formation of a two-dimensional hole gas in the potential well at the diamond surface [Fig. \ref{fig:fig2}(c)], making it conductive. This effect has been used for the development of different types of field-effect transistors (FETs) (solution-gated FET \cite{Kawarada2001}, ion-sensitive FET \cite{Haertl2007}, enzyme-modified FET \cite{Hartl2008}, in-plane gate FET \cite{Garrido2003}), where the number of holes in the surface conductive channel can be modulated over several orders of magnitude with an external gate electrode. The band bending not only leads to accumulation of holes but is also expected to influence the charge state of NV centers. The upwards band bending is likely to render the NV centres positively charged, and thus modifying their fluorescence properties. In this work, we investigate the effect of the surface termination of diamond on the charge state of NV centers using confocal fluorescence microscopy.  \\
Experiments were conducted on electronic grade, synthetic single crystalline diamonds with a nitrogen concentration below 5 ppb (purchased from Element Six Ltd.). NV centers were created by focused N ion beam implantation at energies ranging from 1.5 keV to 30 keV, followed by thermal annealing at 800$^{\circ}$C. The diamond surfaces were thoroughly cleaned in sulfuric acid and oxygen plasma. Samples were fully hydrogen-terminated in a microwave-assisted plasma reactor \cite{Dankerl2009}. A grid-pattern of photoresist was created by conventional photolithography. Subsequent exposure to oxygen plasma selectively oxidized the diamond surfaces.\\ 
Measurements were performed with a home-built confocal microscope. A diode-pumped frequency-doubled Nd:YAG laser at a wavelength of $\lambda = 532$ nm was used to excite the NV centers. The fluorescent signal was recorded by an avalanche photodiode or, alternatively, routed to a spectrometer. Different filter and bandpass combinations were employed to distinguish between NV$^0$ and NV$^-$.\\
\begin{figure}
	\centering
		\includegraphics{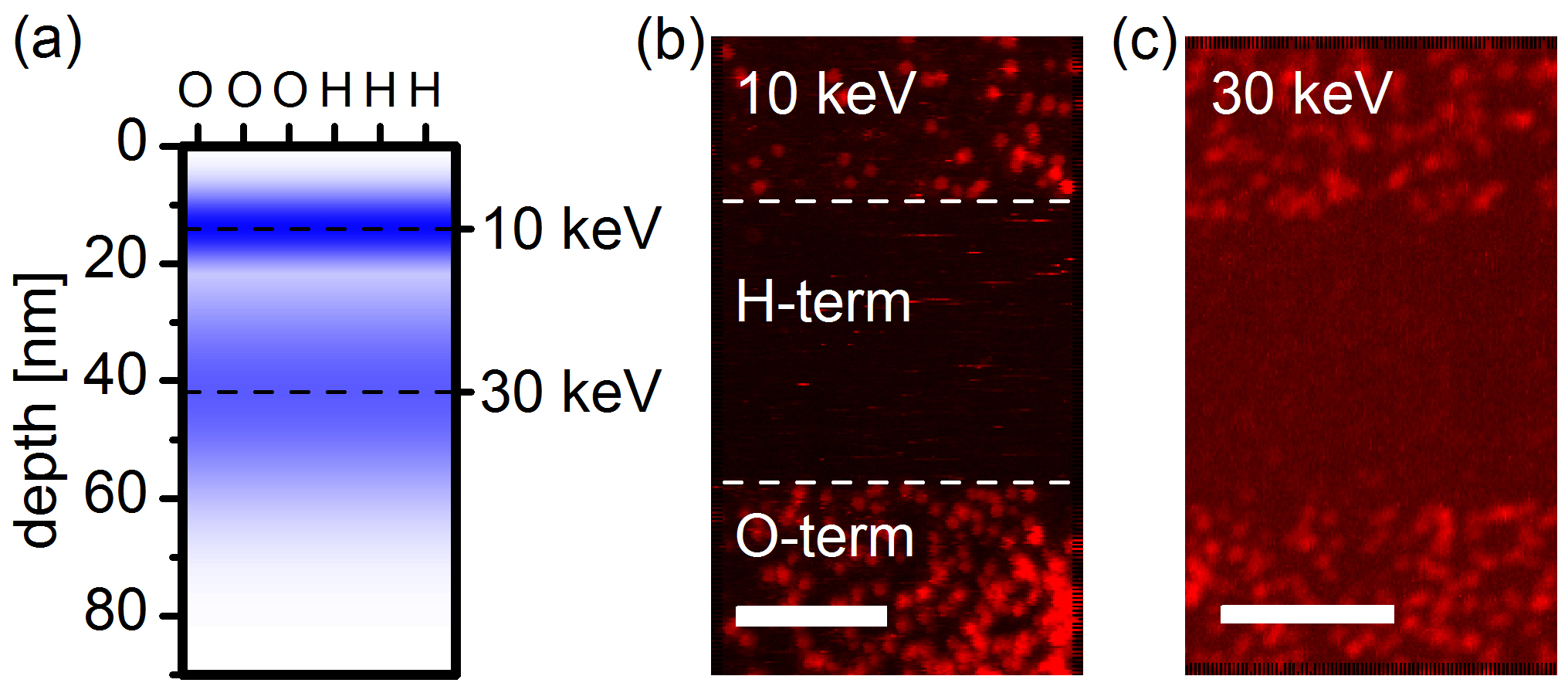}
		\caption{(a) Schematic of the implantation profile at the diamond surface for energies of 10 keV and 30 keV. (b),(c) Confocal scans of diamond surfaces showing the fluorescence intensity of single NV$^-$ (650 nm longpass filter (LP)) for implantation spots of different energies (10 keV and 30 keV). The darker regions correspond to the H-terminated part of the surface. Scale bar is 5 $\mu$m.}
		\label{fig:fig1}
\end{figure}
The effect of surface termination could be confirmed by investigating the fluorescence of NV$^-$ and NV$^0$ in diamonds with different implantation energies, resulting in different implantation depths, and with varying doses. Figure \ref{fig:fig1}(a) shows a schematic representation of the density of implanted NV centers in the first few nanometers at the diamond surface. It should be noted that higher energies do not only lead to a deeper implantation but also to a broader distribution of the NV centers. Figures \ref{fig:fig1}(b), (c) show typical confocal two-dimensional scans across the diamond surface for different implantation energies (10 keV and 30 keV) but with comparable dose ($1 \cdot 10^{10}$ cm$^{-2}$). The O/H surface termination pattern can be clearly recognized, as single negatively charged NVs are observed in the oxygen terminated regions whereas there is very little signal emitted from the H-terminated region. Also, there is no fluorescence from NV$^0$ observed in the H-terminated part. We therefore conclude that the nitrogen-vacancies in this region are neither negatively charged nor neutral but are in an unknown, non-fluorescent state. \\
In the 10 keV sample, a large implantation spot includes a dose gradient from the center to the edge. Figure \ref{fig:fig4} shows scans taken in the center (high dose), outside the center (medium dose), and at the edge (low dose) of the implantation spot. When comparing different doses at the same implantation energy, we observe that at very low doses the contrast between oxygen- and hydrogen-terminated regions is strong. This contrast weakens towards higher doses [Fig. \ref{fig:fig4},(a) - (c)]. Figure \ref{fig:fig4}(d) shows a more quantitative analysis of the dose dependence. Spectra in the hydrogenated and oxygenated regions are evaluated by fitting the spectrum of single NV$^0$ or NV$^-$ (taken from Rondin et al.\cite{Rondin2010}). The dose is inferred from the NV$^-$ population in the oxygen-terminated region, whereas NV$^0$ or NV$^-$ are extracted from the hydrogen-terminated part. It can be seen how the relative amount of NV$^-$ increases from below 10\% at low dose to more than 80\% at high dose. At the same time, the amount of NV$^0$ decreases only slightly from 30\% to below 20\%. The remaining NV centers (rest), which do not show fluorescence are assumed to be positively charged and decrease from more than 60\% at low dose to almost 0\% in the center of the implantation spot.\\
\begin{figure}
	\centering
		\includegraphics{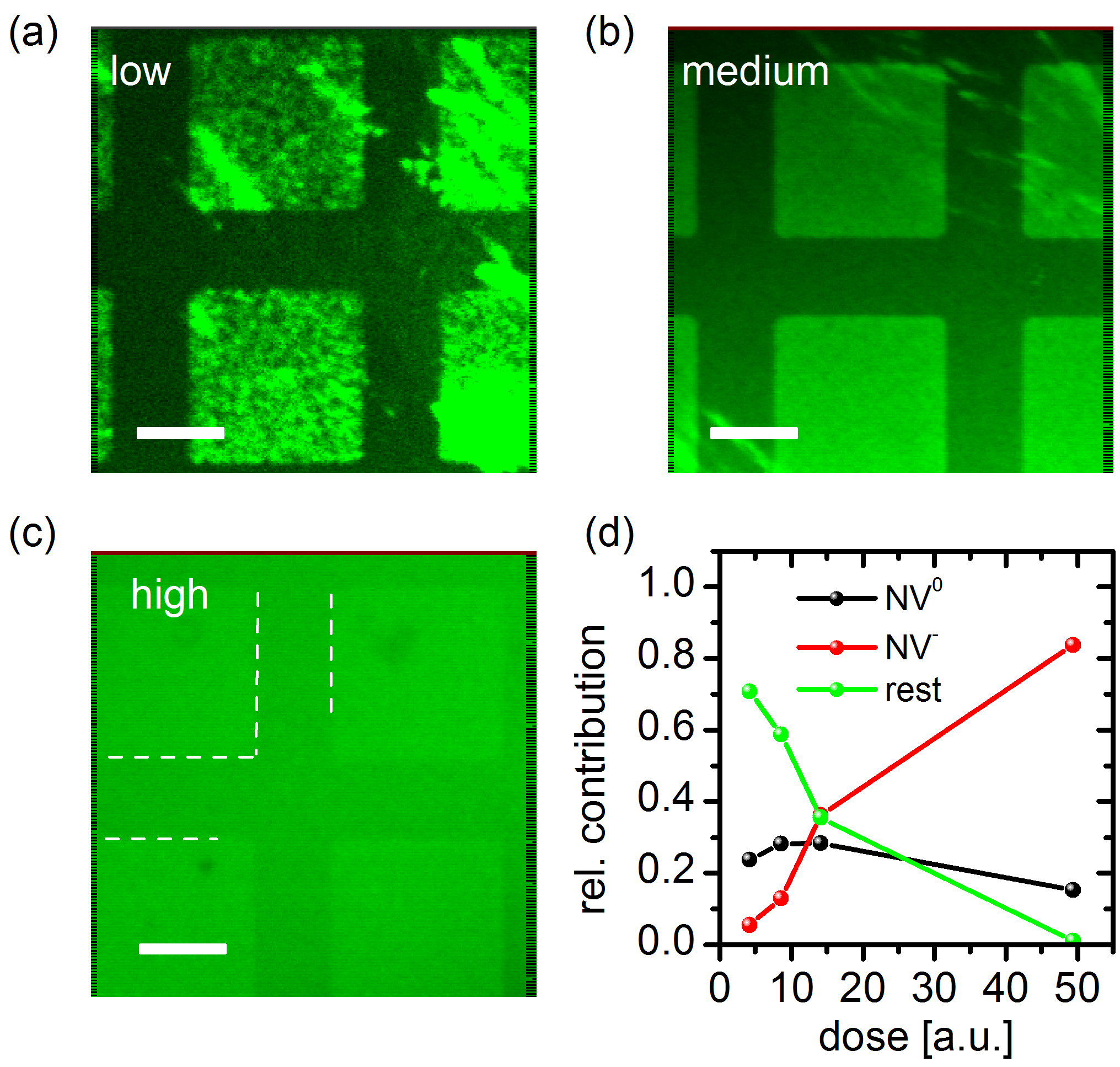}
		\caption{(a)-(c) Confocal scans of a 10 keV implanted diamond surface showing the fluorescence intensity of NV$^-$ (650 nm LP, 532 nm excitation @ 240 µW) at different implantation doses (low, medium, high). In (c) dashed lines indicate the O- and H-terminated regions. Scale bar is 10 $\mu$m. (d) Relative contribution to the local fluorescence intensity of NV$^-$ and NV$^0$, obtained from fits to spectra of single NV$^-$/NV$^0$. }
		\label{fig:fig4}
\end{figure}
In order to get a better understanding of this effect, the band structure of diamond and the charge state of nitrogen-related impurity centers were calculated using the nextnano$^3$ software (www.nextnano.de). Here, the Poisson equation is discretized on a nonuniform grid using the finite differences method and solved iteratively in a self-consistent manner \cite{Birner2007}. The spatial charge $\rho(x)$ and the electrostatic potential $\phi(x)$ distribution are calculated self-consistently by solving the nonlinear Poisson equation
\[\frac{\partial}{\partial x} \epsilon_0 \epsilon \frac{\partial}{\partial x} \phi(x) = -\rho(x) \]
where $\epsilon$ is the static dielectric constant of diamond and $\epsilon_0$ is the permittivity of vacuum. In the diamond, the single band effective mass Schr\"{o}dinger equation is solved and coupled to the Poisson equation via the charge density $\rho(x)$ as described by the wave functions. The occupation of the subbands and defect levels in the band gap is determined by the position relative to the Fermi level. The diamond is modeled in contact with an electrolyte, accounting for the adsorption of a water layer to the hydrophobic surface. A gate voltage $U_G$ is applied between the diamond and an external reference electrode in the electrolyte such that the hole concentration in the surface equals the experimentally determined value for measurements in air. We use the Dirichlet boundary condition $\phi(\infty)=U_G$ for the electrostatic potential in the bulk electrolyte and the Neumann boundary condition $\partial \phi / \partial x = 0$ (vanishing electric field) deep in the diamond. A bulk nitrogen concentration of [N] = 5 ppb is chosen, corresponding to the experimental samples. The n-type substitutional nitrogen is modeled with an ionization energy for N of 1.7 eV \cite{Farrer1969}. As the nitrogen concentration has a strong impact on the position of the bands in the diamond, the additional nitrogen from implantation is taken into account by a Gaussian density profile. The center position and standard deviation of the implanted nitrogen region is related to the implantation energy by SRIM (Stopping and Range of Ions in Matter)-simulations \cite{Ziegler2010}, which give a value of 1.4 nm/keV and 0.46 nm/keV, respectively. Furthermore, nitrogen-vacancies are included in the same region as the implantation of nitrogen at 100 times smaller concentration. The energy level of NV$^{-}$ is set to 2.58 eV below the conduction band minimum \cite{Steeds2000}.\\ 
Figure \ref{fig:fig3}(a) shows the density of negatively charged NVs for a Gaussian implantation profile at depths (center position) of 5 nm (3.6 keV), 10 nm (7.1 keV), and 15 nm (10.7 keV) for an implantation dose of $1 \cdot 10^{10}$ cm$^{-2}$. It can be seen that at an implantation depth of 5 nm almost no NV centers are negatively charged. At a depth of 10 nm already around 37\% of the NVs are negatively charged. The rest of the nitrogen-vacancies is either neutral or positively charged, as the NV$^-$ energy level is already below the Fermi energy. At even deeper implantation (15 nm) the effect of the hydrogen termination becomes smaller and only 24\% of the NVs are not negatively charged. The simulations show that the nitrogen-vacancies close to the surface are in a region where the upwards bending of the energy bands leads to a depletion of electrons in the NV$^-$-level. For a deeper implantation the energy bands drop more slowly to the bulk configuration but as the NV centers are also located deeper in the bulk, more NVs remain neutral or negatively charged. The simulations are performed only for NV$^-$. The addition of the NV$^0$ level would just result in small quantitative but not qualitative changes as the band bending is mainly determined by the concentration of nitrogen in the diamond crystal.
\begin{figure}
	\centering
		\includegraphics{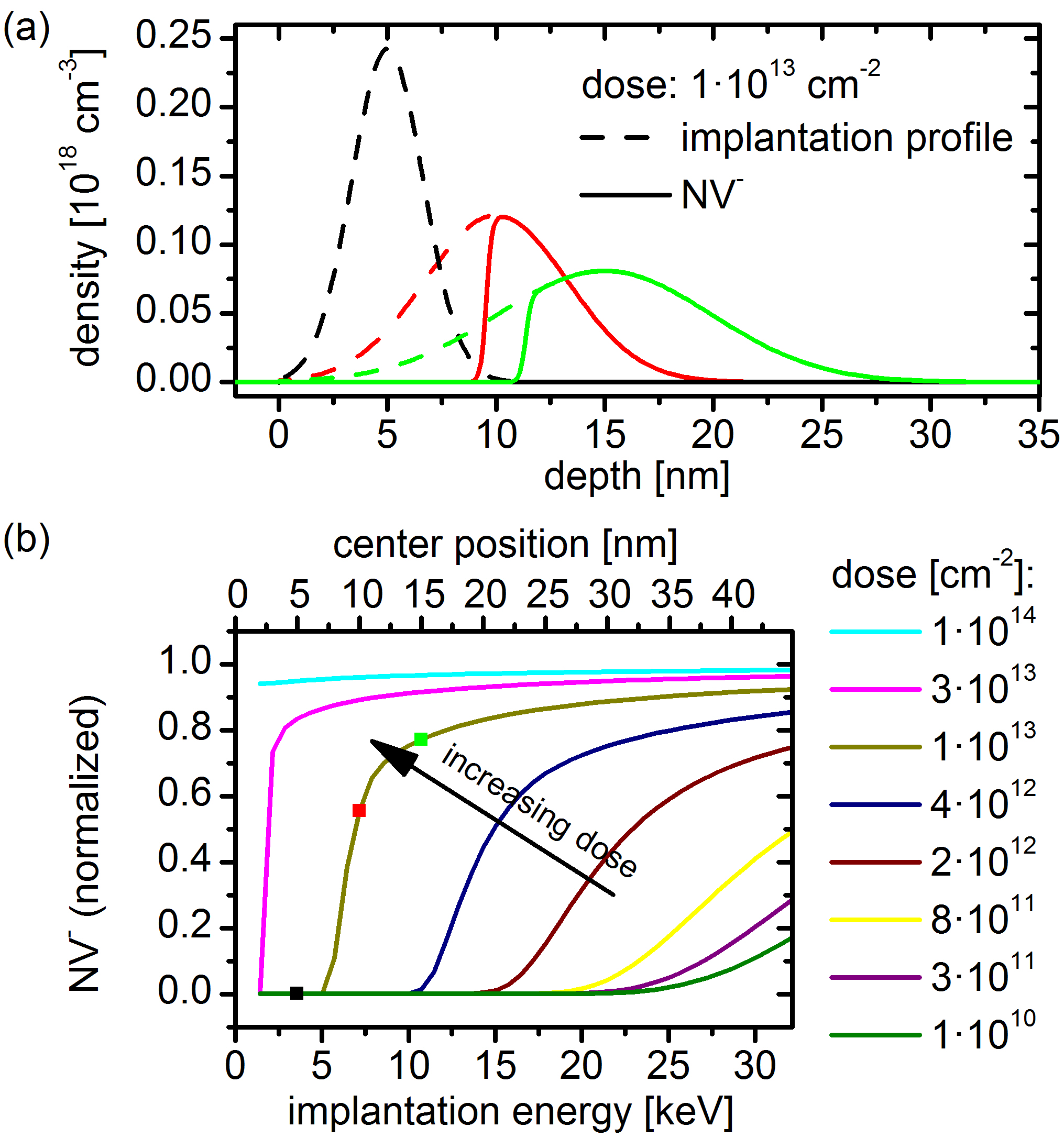}
		\caption{Simulation results from nextnano$^3$. The diamond surface is positioned at 0 nm. (a) density profiles for 3 different implantation energies (3.6 keV, 7.1 keV, and 10.7 keV) at a nitrogen implantation dose of $1 \cdot 10^{13}$ cm$^{-2}$ (dashed line). The solid line represents the density of NV$^-$. (b) Normalized area density of NV$^-$ vs. implantation energy (Gaussian center position) in a range of 1 to 32 keV for different implantation doses.  The solid symbols correspond to the data from (a).}
		\label{fig:fig3}
\end{figure}
The fraction of nitrogen-vacancies which are negatively charged is plotted in Figure \ref{fig:fig3}(b) as a function of the implantation energy for different doses. Again, it can be seen that for low implantation depths and as long as the crystal is not too heavily implanted only a small fraction of NVs are negatively charged, which explains the lack of fluorescence in the H-terminated regions. When implanting deeper in the crystal, more and more negatively charged nitrogen-vacancies contribute to the fluorescence signal as the fraction of NV$^-$ increases. At low nitrogen implantation doses (e.g. $1 \cdot 10^{10}$ cm$^{-2}$) the NVs remain not negatively charged over a large implantation energy range. For higher implantation doses the transition towards negatively charged NVs gets sharper and occurs at lower implantation energies, which can be easily understood when considering how the implantation dose influences the band bending: for a high concentration of implanted nitrogen, the high density of the donor impurities makes the energy bands drop more quickly towards bulk values. Therefore, at the same implantation depth, more NV are negatively charged for higher implantation doses. These findings explain the decrease in contrast between the oxygen- and hydrogen-terminated regions in the 10 keV diamond sample, when moving from the edge of the implantation spot to the center area [Fig. \ref{fig:fig4}(a)-(c)]. In the center of the spot the implantation dose is high, so almost all NV are negatively charged and the brightness in the fluorescence signal of the H-terminated area resembles the one in the O-terminated region. At the outer region a low implantation dose leads to a slow drop of the energy levels and the surface termination strongly suppresses the NV$^-$ fluorescence, even deep in the crystal [Fig. \ref{fig:fig1}(b)].\\
In conclusion, we have demonstrated how the surface termination can be used to control the charge state of the nitrogen-vacancy center in diamond. Compared to an oxygen-terminated surface, the fluorescence of NV$^-$ can be suppressed in a hydrogen-terminated surface due to a decharging or even positive charge of the defect. The deeper the NV centers are implanted into the crystal the weaker an effect of the surface termination is observed. The surface conductivity of hydrogenated diamond can be used to explain this effect, as the accumulation of holes in the H-terminated diamond surface is not compatible with a negatively charged nitrogen-vacancy. Simulations have confirmed this effect, showing how the charge state of the NV depends on the implantation energy. They furthermore reveal the effect of the implantation dose which influences the band bending at the diamond surface, explaining the change in contrast between O- and H-terminated diamond regions with a constant implantation energy but varying dose.\\
The detailed understanding of the effect of the surface termination on the fluorescence of NVs in diamond reported in this work opens the way towards an electrostatic control of the charge state of the NV centers in diamond. An external gate electrode could be used to control the density of charge carriers, and thus the band bending at the diamond surface, and by this provide a way for the switching of single NV centers. 
\acknowledgments
The authors of this publication gratefully acknowledge the contribution of Lucas H. Hess and Stefan Birner from the Walter Schottky Institute and Steffen Steinert and Maximilian Nothaft from the Universit\"{a}t Stuttgart. This work is supported by the Baden-W\"{u}rttemberg Stiftung, the Federal Ministry of Education and Research, BMBF (EPHQUAM), the DFG (SFB/TR21), the Nanosystems Initiative Munich (NIM), and the Graduate School for Complex Interfaces of the Technische Universit\"{a}t M\"{u}nchen.

\end{document}